# Catalyst-Free Growth of Atomically-thin $Bi_2O_2Se$ Nanoribbons for High-performance Electronics and Optoelectronics

Usman Khan,[†] Lei Tang,[†] Baofu Ding,[†] Luo Yuting,[†] Simin Feng,[†] Wenjun Chen,[†] Muhammad Jahangir Khan,[†] Bilu Liu[*,†], and Hui-Ming Cheng[*, †,#]

[†]Shenzhen Geim Graphene Center, Tsinghua-Berkeley Shenzhen Institute & Institute of Materials Research, Tsinghua Shenzhen International Graduate School, Tsinghua University, Shenzhen 518055, P. R. China.

[#]Shenyang National Laboratory for Materials Science, Institute of Metal Research, Chinese Academy of Sciences, Shenyang 110016, P. R. China.

Correspondence should be addressed to bilu.liu@sz.tsinghua.edu.cn (BL), or hmcheng@sz.tsinghua.edu.cn (HMC)

## ABSTRACT

One-dimensional (1D) materials have attracted significant research interest due to their unique quantum confinement effects and edge-related properties. Atomically thin 1D nanoribbon is particularly interesting because it is a valuable platform with physical limits of both thickness and width. Here, we develop a catalyst-free growth method and achieves the growth of $Bi_2O_2Se$ nanostructures with tunable dimensionality. Significantly, $Bi_2O_2Se$ nanoribbons with thickness down to 0.65 nm, corresponding to monolayer, are successfully grown for the first time. Electrical and optoelectronic measurements show that $Bi_2O_2Se$ nanoribbons possess decent performance in terms of mobility, on/off ratio, and photoresponsivity, suggesting their promising for devices. This work not only reports a new method for the growth of atomically thin nanoribbons but also provides a platform to study properties and applications of such nanoribbon materials at thickness limit.

## INTRODUCTION

Dimensionality engineering of materials has been an effective strategy to manipulate their electronic, optical, magnetic, and catalytic properties.[1] One-dimensional (1D) structure is important because of its unique properties including edge states,[2] quantum confinement,[3] ferromagnetic feature,[4] and high integration density for devices.[5] In the past two decades, researchers have shown extensive interests to study 1D materials like nanoribbon, nanotube, nanowire, and nanobelt. Recently, there is a trend to prepare 1D nanoribbon from monolayer two-dimensional (2D) materials to make atomically thin nanoribbons and to explore their properties at physical limits of both thickness and width. For example, graphene nanoribbon possesses sizeable bandgaps which are sensitive to its width, expanding the use of graphene in digital electronics. Regarding 2D transition metal dichalcogenides, researchers have shown that $MoS_2$ nanoribbon is more stable than $MoS_2$ nanocluster and can exhibit intrinsic magnetism under certain edge structures without the introduction of metal doping in nonmagnetic $MoS_2$.[2a] In another material example, it is reported that black phosphorus nanoribbons have a different effective mass of charged carriers and bandgaps, making them promising for optoelectronics and valleytronics.[6] It is clear that modulating the properties of materials is important and dimensionality engineering is an effective strategy.

Among 2D materials, $Bi_2O_2Se$ has been synthesized only recently with exotic properties and emerged as a promising candidate for future electronics and optoelectronics.[7] The 2D $Bi_2O_2Se$ has been incorporated in applications including field-effect transistors (FETs),[8] infrared photodetection,[9] spintronics,[10] photothermal therapy and photoacoustic applications.[11] Similar to other 2D materials, the growth of atomically thin $Bi_2O_2Se$ nanoribbons is highly motivated. However, there is very rare research in growing $Bi_2O_2Se$ nanoribbons. Recently, Tan et. al. developed a bismuth-catalyzed vapor-liquid-solid (VLS) mechanism to grow $Bi_2O_2Se$ nanoribbons and studied their use in FET.[12] The thickness of the thinnest VLS grown nanoribbon was around5 nm, corresponding to 8 atomic layers. FETs based on these $Bi_2O_2Se$ nanoribbons revealed a decent current on/off ratio of $>10^6$ and a room temperature field-effect mobility of 220 $cm^2v^{-1}s^{-1}$. It is intriguing to explore the growth and properties of monolayer $Bi_2O_2Se$ nanoribbons, which is not reported previously. In addition, catalysts might raise impurities to $Bi_2O_2Se$ nanoribbons during the VLS growth process. The metal contamination may introduce energy levels in the bandgap of semiconductors and, influence their optical and electrical performance.[13] Therefore, the development of catalyst-free growth method and exploiting properties of $Bi_2O_2Se$ nanoribbons down to one monolayer are essential but remain challenging.

Here, we developed a catalyst-free CVD method for the synthesis of $Bi_2O_2Se$ nanostructures with tunable dimensionality. By controlling the precursor ratio of $Bi_2O_3$ and $Bi_2Se_3$ in the furnace, the growth of 2D $Bi_2O_2Se$ quadrilaterals can be precisely controlled and consequently $Bi_2O_2Se$ nanoribbons with thickness down to monolayer are grown. The CVD grown $Bi_2O_2Se$ nanoribbon-based FET exhibited a high current on/off ratio of $>10^7$, and high electron mobility of ~262 $cm^2V^{-1}s^{-1}$. Moreover, photodetectors of $Bi_2O_2Se$ nanoribbons show a high photoresponsivity of ~$9.2 \times 10^6$ $AW^{-1}$. The successful growth of $Bi_2O_2Se$ nanoribbons with atomically-thin thickness and decent device performance suggest its great potential for electronic and optoelectronic applications.

## RESULTS AND DISCUSSION

$Bi_2O_2Se$ nanostructures are grown by CVD on a mica ($KMg_3AlSi_3O_{10}F_2$) substrate. The growth of $Bi_2O_2Se$ nanostructures on mica is based on electrostatic interaction between $K^+$ layer of mica and $Se^{2-}$ layer of $Bi_2O_2Se$ as shown in Figure 1a. Combining the electrostatic interaction between mica and $Bi_2O_2Se$ facilitates the lateral growth of nanostructures. Atomically thin growth of 2D $Bi_2O_2Se$ results because of the lower energy barrier. Dual precursors $Bi_2O_3$ and $Bi_2Se_3$ were utilized as co-evaporation sources for the CVD growth of $Bi_2O_2Se$. The experimental setup for the growth of $Bi_2O_2Se$ nanostructures is illustrated in Figure 1b. The typical square-shaped optical microscope (OM) image of $Bi_2O_2Se$ is shown in Figure 1c. The 2D $Bi_2O_2Se$ nanoplate has an ultrasmooth large single crystal with domain size in the range of millimeters. Notably, $Bi_2O_2Se$ with a large domain size of ~ 1.5 mm were obtained by precisely controlling the growth parameters including growth time, growth temperature, and gas flow rate. It was observed that the growth temperature and weight ratio of precursors had a strong impact on the morphology of $Bi_2O_2Se$. Therefore, more precise experiments were conducted by tuning the growth temperature concerning the weight ratio of precursor powders to achieve atomically thin $Bi_2O_2Se$ nanostructures. It was predicted that the synthesis of $Bi_2O_2Se$ nanoribbon went through different growth stages and details are given in supplementary information. As shown in Figure 1d, the as-grown flakes have nanoribbon-like-morphology with an average width and length ~80 nm and ~200 µm, respectively. The high-magnification OM image of $Bi_2O_2Se$ nanoribbons is shown in Figure 1e. It has been observed that the weight ratio of precursors and growth temperature play important roles in achieving nanostructures with multi-dimensionality including square-shaped $Bi_2O_2Se$, rectangular-shaped-$Bi_2O_2Se$, and ribbon-like morphology of $Bi_2O_2Se$ as summarized in Figure 1f. It is worth mentioning that the nanoribbons were prepared without the use of any catalysts.

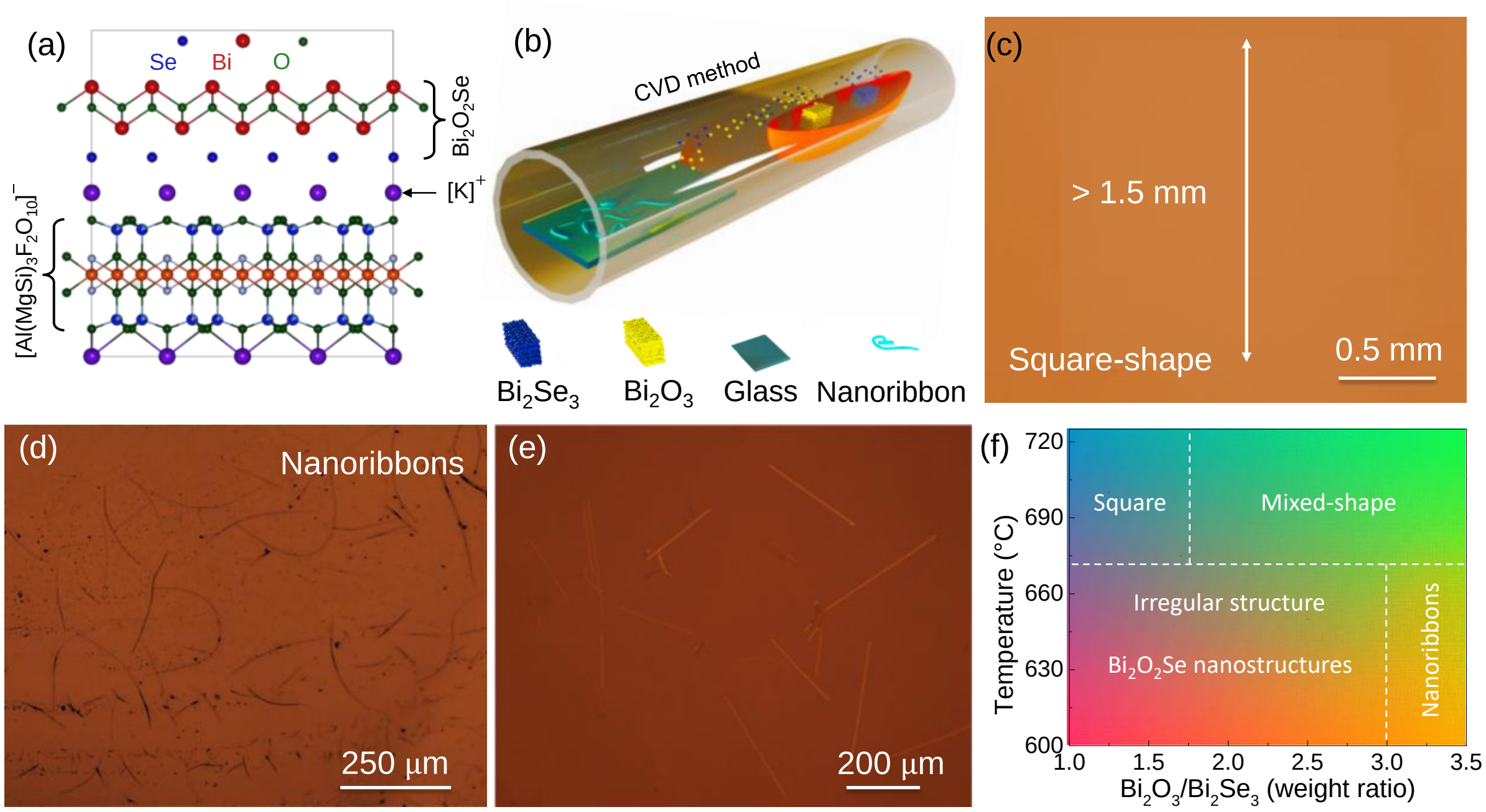

Figure 1: Catalyst-free CVD for the growth of $Bi_2O_2Se$ nanostructures with tunnable dimensionality. a) Schematic illustration of the planar growth of $Bi_2O_2Se$ on mica. b) Schematic illustration for CVD growth of $Bi_2O_2Se$ nanostructures with tunable morphology such as 2D quadrilateral $Bi_2O_2Se$ and 1D $Bi_2O_2Se$ nanoribbons whereas dual precursors of $Bi_2O_3$ and $Bi_2Se_3$ are used in CVD growth. c) A typical OM image of square-shaped $Bi_2O_2Se$ with an average domain size of > 1.5 mm. OM images of $Bi_2O_2Se$ nanoribbons of (d) low resolution and (e) high-resolution image. f) The schematic design of the morphology control of $Bi_2O_2Se$ nanostructures via tuning temperature and the weight ratio of $Bi_2O_3$ and $Bi_2Se_3$ precursors.

Next, we focus on the topography and crystal structure of $Bi_2O_2Se$ nanoribbons. The typical atomic force microscopy (AFM) image of $Bi_2O_2Se$ nanoribbon is displayed in Figure 2a and shows that the topography of the nanoribbon is clean and homogeneous. The thickness of the sample is ~0.65 nm corresponds to the thickness of the monolayer with a width of ~80 nm. Multilayer $Bi_2O_2Se$ nanoribbons have also been synthesized by the CVD method, one sample of which has a thickness of ~12 nm as shown in Figure 2b. Furthermore, Raman spectroscopy of CVD grown multilayer $Bi_2O_2Se$ nanoribbon is conducted at white circular areas highlighted in the inset of Figure 2c. The point spectra at various positions have analogous characteristic $A_{1g}$ peaks which are centered at ~ 159 $cm^{-1}$. In addition, Raman mapping of the characteristic peak $A_{1g}$ of nanoribbon depicts similar color contrast across the whole region, indicating the homogeneity of the sample as shown in Figure 2d. Next, we examined the quality of CVD grown samples based on the chemical composition and crystal structure of $Bi_2O_2Se$ nanoribbons. X-ray photoelectron spectroscopy (XPS) confirms the chemical bonding states of Bi, Se and O as shown in Figure 2e-g. The quantitative investigation indicates Bi to Se atomic ratio of 2:1, demonstrating the formation of nanoribbons with adequate stoichiometry. The XPS spectrum of Bi is fitted with two peaks, centered at ~164 and ~159 eV which are corresponding peaks of metallic Bi in bismuth oxide. The fitted peaks of Se $3d_{3/2}$ and Se $3d_{5/2}$ are centered at 54 and 53.1 eV. All the identified peaks have their binding energies consistent with the composition of $Bi_2O_2Se$ nanoribbons. This attributes that the excess amount of $Bi_2O_3$ in synthesized $Bi_2O_2Se$ nanoribbons samples does not affect the elemental composition and reveals consistency with the published work.[14] Transmission electron microscope (TEM) was employed to further investigate the lattice parameter and crystal structure of $Bi_2O_2Se$ nanoribbons. Figure 2h depicts a typically transferred nanoribbon on Cu grid without any damage. A high-resolution TEM (HRTEM) image of $Bi_2O_2Se$ nanoribbon indicates a d-spacing of 2.7 Å which is accordant with (110) plane of $Bi_2O_2Se$ (Figure 2i).[14-15] The single crystal behavior of the nanoribbon was further analyzed by a selected area electron diffraction (SAED) image that confirms high crystal quality. The bright spots in the pattern (200), $(1\bar{1}0)$, and (110) indicate the tetragonal structure of nanoribbon as shown in Figure 2j. Overall, the analyses indicate that CVD-grown $Bi_2O_2Se$ nanoribbon has good uniformity, stoichiometry, and high quality.

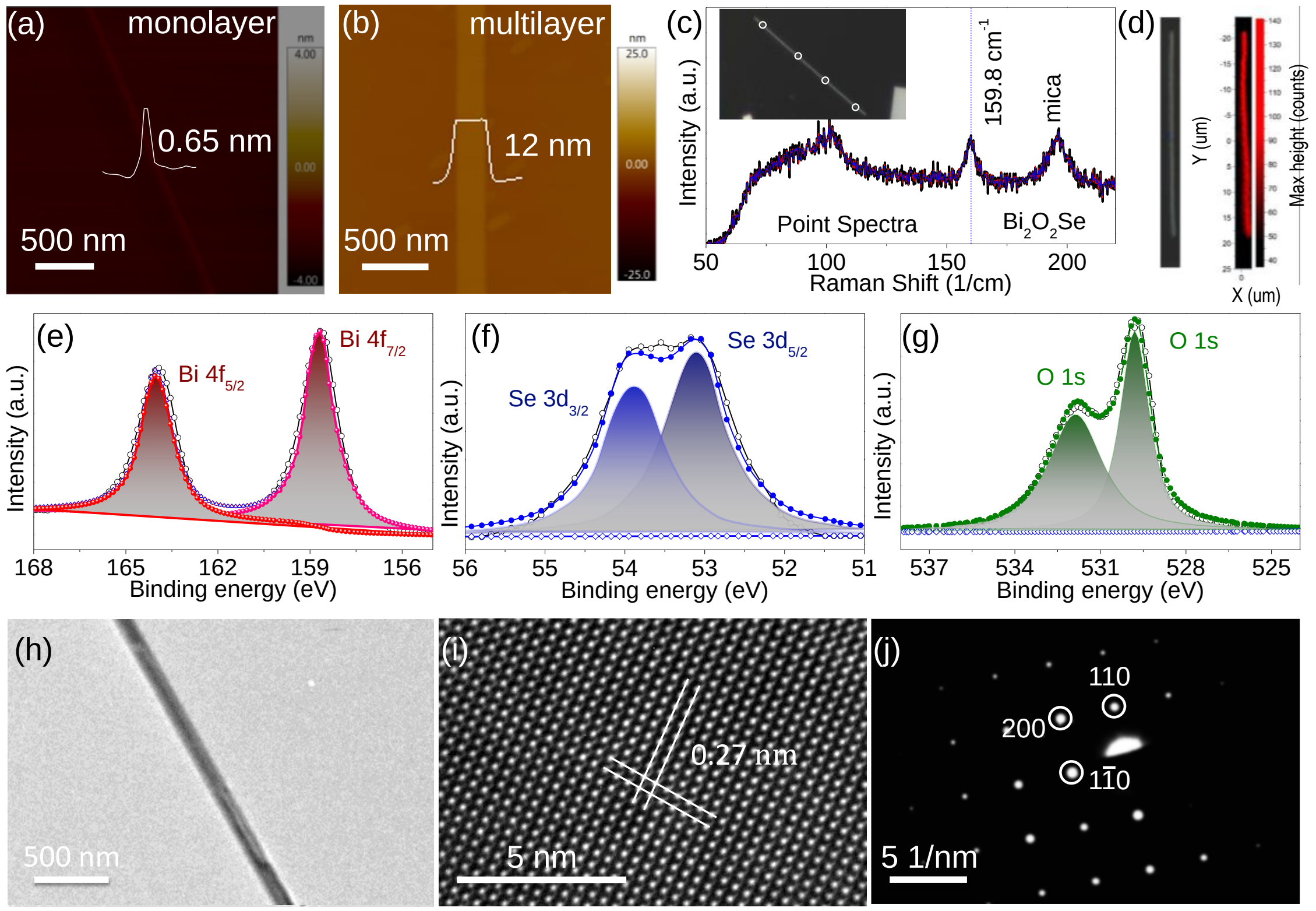

Figure 2: Characterization of monolayer and few-layer $Bi_2O_2Se$ nanoribbons. a) AFM image of a monolayer $Bi_2O_2Se$ nanoribbon with a thickness of ~ 0.65 nm and a width of ~80 nm. b) AFM image of a multilayer $Bi_2O_2Se$ nanoribbon with a thickness of ~12 nm. c) Point spectra of $Bi_2O_2Se$ nanoribbon collected at different points indicating by white circles in the inset. d) Raman mapping of the nanoribbon recorded at $A_{1g}$ vibrational mode. e-g) XPS spectra of the few-layer $Bi_2O_2Se$ nanoribbons. h) TEM image of $Bi_2O_2Se$ nanoribbon. i) HRTEM image of nanoribbon indicating lattice spacing of 0.27 nm for (110) plane. j) A SAED pattern of $Bi_2O_2Se$ nanoribbon representing its single crystalline feature.

Previous studies have shown that atomically thin 2D $Bi_2O_2Se$ nanoflakes represent high electronic response which further motivates the fabrication of FET based on $Bi_2O_2Se$ nanoribbons. Therefore, considering the importance of $Bi_2O_2Se$ nanoribbons for electronic and optoelectronic applications, samples were initially transferred on $SiO_2$/Si substrate. To investigate the electrical performance of $Bi_2O_2Se$ nanoribbons, the transport measurements were investigated with a back-gated FET configuration to determine the feature of thick nanoribbon (12 nm) and studied the contact behavior among nanoribbon and electrodes. The schematic illustration of $Bi_2O_2Se$ nanoribbon-based FET is shown in Figure 3a. The OM image of nanoribbon-based $Bi_2O_2Se$ FET with channel width ($W$) of ~ 780 nm and length ($L$) of ~ 2.1 μm is shown in the inset of Figure 3b. Output characteristic curves ($I_D$-$V_D$) of the device measured at room temperature are represented in Figure 3b. The back-gate voltage ($V_G$) has been tuned from -30 V to 30 V with a step voltage of 10 V. It can be observed that drain current ($I_D$) is highly dependent on $V_G$ which suggests desired gate controllability of the device. The non-linear nature of $I_D$-$V_D$ curves describes the existence of Schottky junction with a slight difference in barrier height among $Bi_2O_2Se$ nanoribbon and metallic electrodes at both ends of the nanoribbon. The $I_D$-$V_D$ curves of nanoribbon FET show n-type semiconducting behavior in which $I_D$ increases (decreases) with a raise in positive (negative) $V_G$. The current ratio amongst $V_G$ of -30 V and 30 V is ~$10^7$, which is depicted from the semi-log scale drawn in Figure 3c. The room temperature transfer characteristic curves ($I_D$-$V_G$) at different $V_D$ are shown in Figure 3d. The $I_D$-$V_G$ curves delve the increase in $I_D$ with $V_G$ which further approves n-type conductivity of nanoribbons. The calculated field-effect carrier mobility is ~262 $cm^2V^{-1}s^{-1}$ and the corresponding carrier density of $Bi_2O_2Se$ nanoribbon FET is $3.2 \times 10^{-12}$ $cm^{-2}$ at which $V_D$ = 5V. The threshold voltage is determined to be about -21 V by extrapolation of $I_D$-$V_G$ curves in the linear region. Figure 3e depicts $I_D$-$V_G$ curves in linear scale at different $V_D$ representing the n-type feature of FET device. The feature is more prominent in the logarithmic scale of $I_D$-$V_G$ curves (Figure 3f). Table 1 benchmarks the typical research on nanoribbon designed by various synthesis techniques, materials, and electrical performance. We summarized some well-known research on nanoribbon-based FET such as n-type TMDCs, including $MoS_2$,[16] CdSe,[17] tri-chalcogenides $TiS_3$,[18] p-type graphene nanoribbon,[19] superlattices of AlGaN/GaN,[20] and III–V compound like boron nitride.[21] The FET based on $Bi_2O_2Se$ nanoribbons not only achieves high field-effect mobility in comparison with graphene nanoribbon but also simplifies the assembly complexity required for the opening of graphene bandgap. Overall, these experimental results show that $Bi_2O_2Se$ nanoribbon has a high-quality feature and is a promising candidate for high-performance electronics.

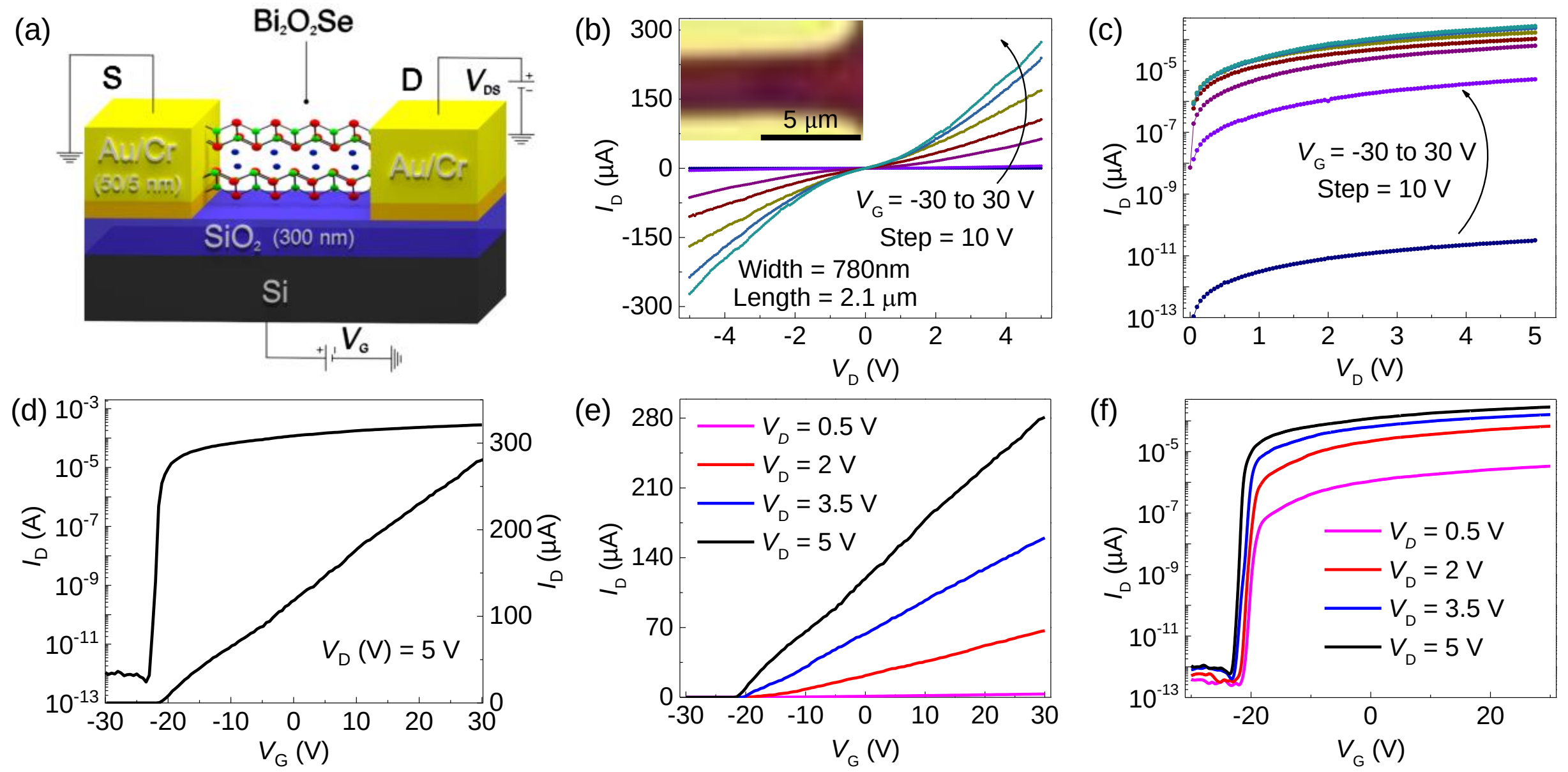

Figure 3: Electrical performance of 1D $Bi_2O_2Se$ nanoribbon. a) Schematic illustration of back-gated $Bi_2O_2Se$ nanoribbon-based FET. b) The inset depicts OM of $Bi_2O_2Se$ nanoribbon-based FET device with a channel length ($L$) and width ($W$) of 2.1 μm and 780 nm, respectively, whereas the main figure indicates output characteristic curves ($I_D$-$V_D$) of the device as a function of gate voltage ($V_G$) in linear scale (b) and logarithmic scale (c). d) The transfer characteristic curves ($I_D$-$V_G$) at $V_D$ of 5V. e-f) $I_D$-$V_G$ family curves of the device as a function of $V_D$ are represented in linear and logarithmic scales.

**Table 1.** Performance comparison of nanoribbon field-effect transistors reported in the literature.

| Materials | Method | Nature (-type) | FET architecture | On/off ratio | Mobility ($cm^2/Vs$) | Reference |
|---|---|---|---|---|---|---|
| Black phosphorus | Plasma etching | p | Top-gate | $10^3$ | 310 | [22] |
| Graphene | Unzip MWCNTs | p | Back-gate | - | 1500 | [19a] |
| Graphene | Bottom-up approach | p | Back-gate | $10^6$ | 200 | [19b] |
| hBN | MWBNNTs | n | - | - | 58.8 | [21] |
| AlGaN/GaN | Plasma etching | n | Top-gate | - | 289 | [20] |
| $TiS_3$ | Heat treatment | n | Back-gate | $10^4$ | 2.6 | [18] |
| CdSe | Thermal Evaporation | n | Back-gate | $10^4$ | 9.6 | [17] |
| $MoS_2$ | Plasma etching | n | Back-gate | $10^7$ | 21.8 | [16] |
| $Bi_2O_2Se$ | VLS | n | Top-gate | $10^6$ | 220 | [12] |
| $Bi_2O_2Se$ | CVD | n | Back-gate | $>10^7$ | ~262 | This work |

Notes. MWCNTs=Multiwall carbon nanotubes, MWBNNTs=Multiwall boron nitride nanotubes.

Generally, few-layer semiconductors with suitable bandgaps are promising materials for photodetectors because of low energy loss during charge recombination, enhanced light absorption and better drain to source charge transfer.[23] Considering the advantage of the suitable bandgap of $Bi_2O_2Se$, we characterized the performance of photodetector based on $Bi_2O_2Se$ nanoribbon and is illustrated in Figure 4a. Figure 4b reveals zero-gated $I_D$-$V_D$ curves in dark and under light illumination. The $I_D$ of the device on a linear scale at different power densities is represented in the inset of Figure 4b. The device exhibits dark current ($I_D$) of ~100 μA at $V_D$ = ±5 V. The maximum photocurrent ($I_{ph} = I_{light} – I_{dark}$) generated at $V_D$ = 5 V ($V_G$ = 0 V) is 130 μA under power density of 66.84 W/$m^2$. The photocurrent ($I_{ph}$) increases under illumination because of increased photogenerated charge density. The low power of laser light would possibly excite the electron-hole pairs in the nanoribbon. These pairs diffuse to the depletion region, where the built-in electric field is strong enough to dissociate these pairs into the free electrons and holes. The collection of these free charges by electrodes finally results in the linear increase of photocurrent. Under strong light intensity, the newly charge-balanced state stimulates incoming charge carriers at Schottky junction and depletion region squeezes. The depletion width is consequently reduced, thereby decreasing the strength of the electric field and the resultant dissociation efficiency of electron-hole pairs. This hinders the further increase of photoresponse,[24] as presented by the saturation behavior of photocurrent ~ 66.84 $W/m^2$ in Figure 4b. The trend of $I_{ph}$ as a function of different light intensities follows the power-law $I_{ph} \propto P^{\gamma}$ where γ represents $I_{ph}$ response on light and is shown in Figure 4c. The values of the exponential function γ have been derived by a non-linear curve fitting and it yields γ = 0.55, 0.5, 0.44, and 0.39 for $V_D$ of 5, 4, 3, and 2 V, respectively. The non-unity of γ suggests complex mechanisms that would involve charge trapping, electron-hole pair generation and recombination in $Bi_2O_2Se$ nanoribbons.[25] The time-resolved photoresponse of $Bi_2O_2Se$ nanoribbon-based photodetector is measured by periodically switching of 590 nm laser light with a light intensity of 66.84 $W/m^2$ at $V_D$=5 V (Figure 4d). It can be depicted that $I_D$ increases abruptly and becomes stabilized at ≈230 µA, followed by dropping down to initial $I_D$ ≈ 100 µA after removal of light. The device responded excellent on and off swapping of voltage with good stability and reproducibility which depicts excellent reproducibility of photodetector. The rise ($\tau_{rise}$) and decay ($\tau_{decay}$) time constants of the device are obtained with normalized temporal photoresponse as shown in Figure 4e. By defining $\tau_{rise}$ ($\tau_{decay}$) as the time required by the photodetector to increase (decrease) photocurrent from 10% $I_D$ to 90% $I_D$ (90% $I_D$ to 10% $I_D$), it is characterized to be ~39 ms (63 ms).

The other figure of merits of the photodetectors are responsivity (*R*) and detectivity (*D**) and are represented in Figure 4f. The maximum spectral *R* of the $Bi_2O_2Se$ photodetector has been measured to be 9.19 × $10^6$ $AW^{-1}$ at zero-gate voltage. The high value of *R* is because of a strong induced electric field at $Bi_2O_2Se$ nanoribbon and metal contact Schottky junction. The corresponding value of specific detectivity at zero-gate voltage is 2.08 × $10^{12}$ Jones. The outstanding performance of the device is because of the facts: (i) single-crystalline nature of nanoribbons that suppresses scattering centers and attributes more efficient carriers at contact region and (ii) absence of contamination that can deteriorate the carrier traps and reduce the non-radiative recombination centers in nanoribbons. The *R* as a function of light intensity is shown in Figure 4f. It is observed that the *R* decreases as the light intensity rises which can be elucidated as follows: three different mechanisms occur due to the irradiation of light on semiconducting nanoribbons, namely electron and hole pair generation, electron and hole pair recombination and transport of photogenerated charges due to a strong induced electric field.[25b, 26] The photogenerated charge carriers are proportional to the intensity of incident light. The strong induced field suppresses the recombination at low intensity and generates high *R*. Overall, these experimental results together with tunable dimensional growth of $Bi_2O_2Se$ depict that $Bi_2O_2Se$ nanoribbon is a promising candidate for high-performance optoelectronics.

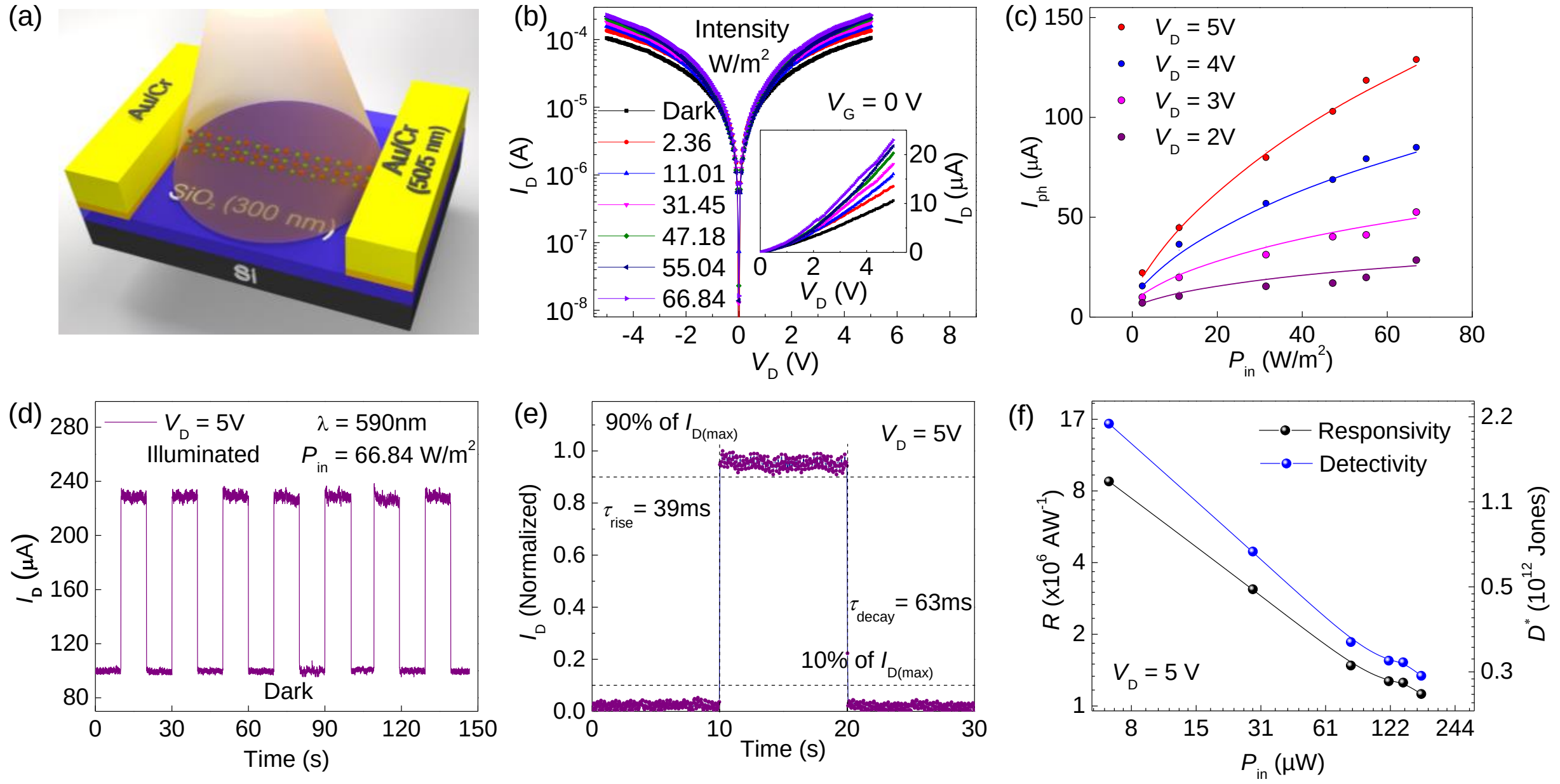

Figure 4: Optoelectronic performance of 1D $Bi_2O_2Se$ nanoribbon. a) The output characteristic curves ($I_D$-$V_D$) of zero-gated $Bi_2O_2Se$ nanoribbon photodetector under light illumination of 590 nm at different intensities; the linear scale of curves is shown in the inset. b) The photocurrent variation with light intensity at zero-gate bias for different drain voltages follows the power law. c) The on to off photocurrent to dark current ratio for various light intensities at zero-gate voltage.

d) The on/off current switching of $Bi_2O_2Se$ nanoribbon photodetector for laser intensity of 66.84 $W/m^2$ depicts repeatability and stability of the device. e) Time-resolved photoresponse of the photodetector reveals the rise and decay times of 39 and 63 ms, respectively. f) Variation of spectral responsitivity and specific detectivity of the device as a function of incident power drain voltage.

## CONCLUSION

We have developed a catalyst-free CVD method to grow $Bi_2O_2Se$ with tunable dimensionality ranging from 2D quadrilateral shape to 1D nanoribbons on a mica substrate. The morphology of nanostructures was controlled by tuning the growth conditions and mechanisms. Monolayer $Bi_2O_2Se$ nanoribbons with a thickness of 0.65 nm were grown for the first time. The FETs made of $Bi_2O_2Se$ nanoribbons exhibited n-type semiconductor characteristics with high electron mobility of ~262 $cm^2V^{-1}s^{-1}$ and a high on/off ratio of $>10^7$. In addition, $Bi_2O_2Se$ nanoribbon-based photodetector exhibits decent photoresponsivity of ~ $9.2 \times 10^6$ $AW^{-1}$. Considering their high quality, good uniformity and stoichiometry, our results suggest the potential of these monolayer thick nanoribbons for a wide range of applications.

## AUTHOR INFORMATION

Corresponding Author

*E-mail: bilu.liu@sz.tsinghua.edu.cn, hmcheng@sz.tsinghua.edu.cn

Notes

The authors declare no competing financial interest.

## ACKNOWLEDGMENTS

We acknowledge support by the National Natural Science Foundation of China (Nos. 51991340, 51920105002, 51991343, and 51950410577), Guangdong Innovative and Entrepreneurial Research Team Program (No. 2017ZT07C341), the Bureau of Industry and Information Technology of Shenzhen for the "2017 Graphene Manufacturing Innovation Center Project" (No. 201901171523), and the Shenzhen Basic Research Project (Nos. JCYJ20200109144620815, JCYJ20190809180605522 and JCYJ20200109144616617).